\documentstyle[aps,twocolumn,prl,tighten]{revtex}

\input epsf
\def\plotone#1{\centering \leavevmode
\epsfxsize= 1.0\columnwidth \epsfbox{#1}}

\def\apjl{Astrophys. J. Lett. }
\def\mnras{Mon. Not. R. astron. Soc. }
\def\araa{Annu. Rev. Astron. Astrophys. }
\def\aj{Astron. J. }

\def\be{\begin{equation}}
\def\ee{\end{equation}}
\def\bea{\begin{eqnarray}}
\def\eea{\end{eqnarray}}

\def\gev{\,{\rm GeV}}

\def\rcm{\,{\rm cm}}
\def\pc{\,{\rm pc}}
\def\kpc{\,{\rm kpc}}

\def\ev{{\,\rm eV}}

\def\cmm2{{\,\rm cm^{-2}}}
\def\cm2{{\,{\rm cm}^2}}
\def\cmm3{{\,{\rm cm}^{-3}}}
\def\gcmm3{{\,{\rm g\,cm^{-3}}}}
\def\kms{\,{\rm km\,s^{-1}}}

\def\fun#1#2{\lower3.6pt\vbox{\baselineskip0pt\lineskip.9pt
  \ialign{$\mathsurround=0pt#1\hfil##\hfil$\crcr#2\crcr\sim\crcr}}}

\def\hyi{H\thinspace{$\scriptstyle{\rm I}$}~}

\def\kms{{\rm~km~s^{-1}}}

\def\kpc{{\rm~kpc}}

\def\msun{{\,M_\odot}}

\def\eg{{e.g., }}

\def\etal{{\it et al.}}

\def\p3m{P$^3$M}

\def\la{\mathrel{\mathpalette\fun <}}
\def\ga{\mathrel{\mathpalette\fun >}}
\def\fun#1#2{\lower3.6pt\vbox{\baselineskip0pt\lineskip.9pt
  \ialign{$\mathsurround=0pt#1\hfil##\hfil$\crcr#2\crcr\sim\crcr}}}

\newcommand{\order}[1]{{\cal O}(#1)}

\begin{document}
\twocolumn[\hsize\textwidth\columnwidth\hsize\csname @twocolumnfalse\endcsname
\draft
\title{Annihilating Cold Dark Matter}
\author{Manoj\ Kaplinghat,$^{1}$ Lloyd\ Knox,$^{1}$ and
Michael\ S.\ Turner$^{1,2,3}$}
\address{$^1$ Department of Astronomy and Astrophysics\\
The University of Chicago, 5640 S. Ellis Ave., Chicago, IL 60637, USA}
\address{$^2$ NASA/Fermilab Astrophysics Center\\
Fermi National Accelerator Laboratory, PO Box 500\\
Batavia, IL  60510-0500, USA}
\address{$^3$ Department of Physics, Enrico Fermi Institute\\
The University of Chicago, Chicago, Illinois 60637, USA}

\date{\today}
\maketitle

\begin{abstract}
Structure formation with cold dark matter (CDM) predicts halos 
with a central density cusp, which are observationally disfavored.
If CDM particles have an annihilation cross section 
$\sigma v \sim 10^{-29} (m/\gev)\rcm^2$, then annihilations 
will soften the cusps. We discuss plausible scenarios for avoiding 
the early Universe annihilation catastrophe that could result from 
such a large cross section.
The predicted scaling of core density with halo mass depends 
upon the velocity dependence of $\sigma v$, and
s-wave annihilation leads to a core density nearly independent of 
halo mass, which seems consistent with observations. 
\end{abstract}
]
{\parindent0pt\it Introduction.}
The idea that the large-scale structure developed by gravitational
instability from initially small-amplitude, adiabatic and nearly
scale-invariant fluctuations
is compatible with a number of observables across a wide range of
length scales (\eg from the cosmic microwave background anisotropy
to the Lyman-$\alpha$ forest).  Essential to this compatibility is the
existence of cold dark matter: matter which is non-baryonic, has only
very weak interactions with photons and baryons, and (prior to gravitational
collapse) is cold.

The greatest challenge
to this otherwise successful scenario comes from the
apparent discrepancy between predicted dark-matter
halo-density profiles and those inferred from observations.
Simulations with non-interacting cold dark matter lead to
halo density profiles that are singular at the 
center \cite{nfw} \cite{singular},
whereas observations indicate uniform density cores.  In this
{\it Letter} we explore the possibility that the dark matter today has
a large cross-section for annihilation which results in preferential
destruction in high-density regions, softening halo cores \cite{mooredidit}.

Detecting and determining the properties of the dark matter
is a major goal of observational cosmology.  If annihilations are
indeed altering the properties of dark matter halos,
then we have a new means of studying the
dark matter.  The interactions of the CDM particles determine
both the magnitude and velocity dependence of the annihilation
cross section.  For example,
for s-wave annihilation, $\sigma_A |v|$ is independent of velocity
and for p-wave annihilation, $\sigma_A |v|$ is proportional to $v^2$.
These two different dependences result in different scaling relations
between core density and halo velocity dispersion, which can
be tested by current observations.

As we show below, current data for high velocity dispersion
systems such as clusters of galaxies to low velocity
dispersion systems such as galactic satellites are
consistent with the same core density of about 
$1 \gev/\rcm^3\, (= 0.026 \msun/\pc^3$) \cite{citem}.
This scale invariance can be explained by s-wave annihilation
with a cross-section $\sigma v \sim 10^{-29} (m/\gev) \rcm^2$,
although future improvements in both the data and predictions
will be necessary before such a statement can be made with
confidence.
As we shall discuss, the cosmological and astrophysical constraints
on annihilating CDM point to a candidate beyond those currently
favored (\eg axion, neutralino).

{\parindent0pt\it Halos of Annihilating Dark Matter.}
Numerical simulations of structure formation in the CDM scenario
show that the dark matter halos which form with a wide range
of masses are all well-fit with the so-called
NFW \cite{nfw} form for the density profile.
This form has $\rho \propto r^{-3}$ at large $r$ and
a cuspy inner region with $\rho \propto r^{-\alpha}$ with
$\alpha=1$.  More recent higher resolution simulations \cite{alpha1.5}
predict cusps that are even stronger, with $\alpha \simeq 1.5$.
Nevertheless, in most of what follows, we use the NFW theory for simplicity.

To be precise, the NFW profile is 
\begin{equation}
\rho(r=xr_s)=\rho_s x^{-1} (1+x)^{-2},
\end{equation}
where the value of $\rho_s$ is determined by the
mean density of the Universe at the time the halo collapsed.  In
CDM theory, small objects collapse first, followed later by larger ones.
Thus, there is an inverse relationship between $\rho_s$ and halo size.
In Fig. 1 we show this scaling relation with halo size
represented by velocity dispersion for the halo,
estimated as $\sigma_{\rm vir}=\sqrt{GM_{\rm vir}/2r_{\rm vir}}$.
(The virial radius, $r_{\rm vir}$, is defined such that the
mean density inside the $r_{\rm vir}$ sphere is 200 times the present
mean density of the Universe, and $M_{\rm vir}$ is the mass contained
within $r_{\rm vir}$ \cite{nfw}.)

Annihilations will alter the halo profiles near the core where
the density of the dark matter particles is the highest.
The annihilation rate (per particle) $\Gamma = n \langle \sigma |v|\rangle$
depends on the velocity dispersion.  We parameterize the velocity
dependence as $\Gamma=(\rho /m)\sigma_A v^n$ ($n=0$ for s-wave; 
$n=2$ for p-wave), where $v$ is the velocity dispersion and $m$ is the 
CDM particle mass.
\begin{figure}[thbp]
\plotone{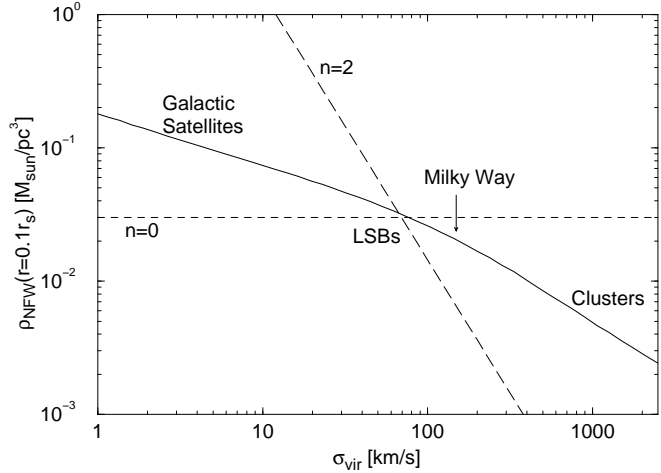}
\caption[caption]{
The halo density at $r = 0.1 r_s$ (where the cusp problem
becomes prominent) in structures of different size according to
NFW \cite{nfw} (solid curve).
The objects are characterized by their virial velocity dispersion
as indicated.  Annihilation lines (dashed curves), normalized to 
LSB galaxies, are shown for the cases $n=0$ and 2.  Above the line, 
annihilations are very important (at $r=0.1r_s$) and below the line 
they are unimportant.
For $n=0$, LSB and smaller objects have their cores softened significantly,
while clusters do not, consistent with observations.  For $n=2$,
clusters would be adversely affected.
}
\label{rho-sigma}
\end{figure}
Fig. \ref{rho-sigma} also shows in a qualitative way how annihilations affect
the core structure of different objects.  The annihilation lines drawn
show whether or not annihilations are important at a NFW halo radius of
$0.1 r_s$ in different kinds of objects.  (Note, since halo densities
diverge, for any object annihilations become significant deep enough
into the core).  The annihilation lines are normalized to soften the
cores of low surface-brightness (LSB) spiral galaxies. 
Because of how annihilations scale with velocity,
for n = 0 clusters remain unaffected at $r\ga 0.1r_s$, while the cores of LSBs
and smaller objects are dramatically softened.  For n = 2, the opposite
is true, which contradicts observations that indicate the NFW profile
works well for clusters.  We expect any $n > 1$ to be inconsistent with
observations.  The case of $n = 0.5$ is interesting since the
annihilation line runs parallel to the structure line
(for $\sigma_{\rm vir} \la$ 100 km/s), 
implying that all the systems will be smoothed off at the same value 
of $r/r_s$.

{\parindent0pt\it Model Building Constraints.}
For the annihilations to be effective in galaxy cores today,
the annihilation rate must satisfy the (approximate) constraint:
\begin{equation}
\Gamma\sim \left(\rho/\rho_{\rm LSB}\right)\,
\left(v/v_{\rm LSB}\right)^n H_0\,,
\label{constraint}
\end{equation}
where the subscript LSB denotes the appropriate values for a
typical LSB and $H_0=100h\kms$ is the present expansion rate of the Universe.
Outside collapsed objects today, the density of CDM is much
lower and annihilations will be unimportant for $n\ge 0$.  
The early Universe is another matter as densities were much
higher, $\rho\propto T^3$, where $T$ is the cosmic background radiation
temperature.

The figure of merit for the effectiveness of annihilations in
the early Universe is measured by annihilation rate divided by 
the expansion rate:  when $\Gamma /H > 1$ annihilations are effective 
(and vice versa).
Assuming that the velocity dispersion of the CDM particles can be
characterized by the background radiation temperature and 
normalizing the cross section to the desired value today, the
temperature dependence of $\Gamma /H$ is
\begin{equation}
\frac{\Gamma}{H} \sim 10^9\left(\frac{T}{\gev}\right)
\left(\frac{T}{10^{-3}m}\right)^n
\sqrt{\frac{T}{T+T_{\rm eq}}},
\label{fom}
\end{equation}
where $T_{\rm eq}\sim 1\ev$ is the temperature at matter -- radiation
equality.  There are three important things to note:  
(1)  the large coefficient
in front of this expression -- annihilations in the early Universe are
a significant consideration; (2) for $n=-1$, the effectiveness of annihilations
is epoch independent and disastrous; and (3) for $n > -1$ annihilations
were more important in the past.

Observational data suggest that if halos are made of annihilating CDM
particles, their annihilation cross section is characterized by
$n\la 1$.  Thus we will focus on $n>-1$, where 
the danger of annihilations is in the past:  $\Gamma /H > 1$ for
\begin{equation}
T > T_A \sim 10^{-3(3+n)/(1+n)}\,{\rm GeV}\, (m/{\rm GeV})^{n/(1+n)}\,,
\end{equation}
or $1\ev$ for $n=0$.  To ensure that early annihilations
do not reduce CDM particles to negligible
numbers, they must be protected against annihilation in
the early Universe.  We suggest two mechanisms; doubtless,
there are other possibilities.

First, CDM particles could be produced late ($T< T_A$)
by the decays of another massive particle.
Note that this requires a long lifetime, $\tau > t(T_A)\sim
10^5\,$yrs, and the mass difference between the two
particles should be small enough to ensure that the relativistic 
decay products do not make the Universe radiation dominated.

The second way of avoiding the early-Universe annihilation catastrophe
is to make the mass of the annihilation product be dynamical.
For example, a phase transition that takes place at $T< T_A$ could
change annihilation from being kinematically impossible to possible if
the mass of the annihilation product dropped below threshold
after the phase transition (or if the mass of the CDM particle rose
above threshold).  A variation on this theme is
coupling the annihilation produced particle to a scalar field, $\phi$, 
with $\langle \phi \rangle \ne 0$.  As $\langle \phi \rangle$ decreases,
either quickly to zero as a result of a symmetry-restoring phase transition,
or slowly as $\langle \phi \rangle$ rolls to the minimum of its potential,
the product particle's mass may drop below threshold, opening
up the new annihilation channel, at $T < T_A$. 

Finally, the CDM annihilation products must not include
photons because their $\gamma$-ray flux would far exceed observational
limits.  For example for 1 GeV CDM particles, the flux would be
around $10^5 \rcm^{-2} {\rm sr}^{-1} {\rm s}^{-1}$,
some ten orders of magnitude above the observed diffuse
$\gamma$-ray flux at 1 GeV.

{\parindent0pt\it Observational Constraints.}
We henceforth restrict ourselves to $n=0$.
The contribution of annihilations to the evolution of the
density profile is given by
\begin{equation}
d \left[\rho(r)/\rho_A\right]/d t=-\left[\rho(r)/\rho_A\right]^2t_0^{-1}\,,
\label{drhodt}
\end{equation}
where $\rho_A\equiv m/(\sigma_A t_0)$ and $t_0$ is the age of
the Universe today. 
Assuming the initial \cite{initial} profile to be NFW, 
the resulting density profile is
\begin{equation}
\rho(r) = \rho_s \left[x(1+x)^2+\rho_s/\rho_{\rm core}\right]^{-1}\,,
\label{n=0}
\end{equation}
where $x\equiv r/r_s$, and a core of constant density, 
$\rho_{\rm core}=\rho_A$, is clearly evident. 
However, the mass loss due to annihilations results in adiabatic 
expansion of the core, such that the quantity $M(r)r$ 
is left invariant \cite{richstone}. 
This expansion results in a lower core density and one can
estimate that the ratio $\rho_{\rm core}/\rho_A$
ranges from about 0.1 (dwarf galaxies) to about 0.3 (clusters).
We have verified this by more detailed numerical work which
allows us to determine $\rho_{\rm core}/\rho_A$ as a function of
halo mass.

\begin{figure}[thbp]
\plotone{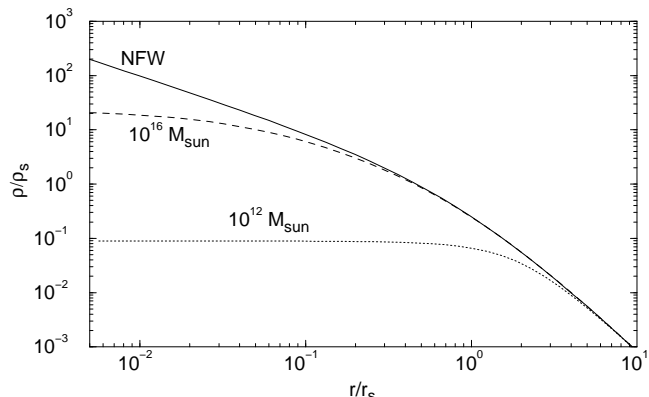}
\caption[caption]{
The solid curve is the NFW density profile. 
Annihilation-modified profiles are labeled by the 
virial mass of the halo: $10^{16} \msun$ (cluster) 
and $10^{12} \msun$ (galaxy).
}
\label{profiles}
\end{figure}

We now turn to the observable constraints on annihilating CDM.
A robust prediction of the s-wave annihilation scenario is that
the cores are more evident in smaller mass halos, 
as can be seen in Fig.~\ref{profiles}. 
So we first turn to the galactic satellites in
the Milky Way group \cite{satellites}, of which there are 11 known.
For a $10^8 \msun$ galactic satellite, the
core radius produced by annihilations is about 1 kpc, which is
about the same as the cut-off radius induced by tidal forces. 
Most of the galactic satellites have large 
velocity dispersions ($\sim$ 10 km/s) for their stellar content, which 
suggests that they are CDM dominated \cite{tides}.
If so, their internal velocity dispersions indicate that
$\rho_{\rm core}= \order{1 \gev/\rcm^3}$ \cite{dSphs-core}.

We also looked at dwarf spiral galaxies and LSBs.
One must use these with caution since
van den Bosch \etal \cite{bosch} have recently claimed that
most of the \hyi rotation curve data do not have sufficient spatial
resolution to put meaningful constraints on the halo cusps.
They do identify three nearby galaxies which
have sufficient spatial resolution -- NGC 247, DDO 154 and 
NGC 3109 \cite{moore}.
van den Bosch \etal \cite{bosch} find that
$0.55 < \alpha < 1.26$ for the LSB (NGC 247),
and $\alpha < 0.5$ for the two dwarfs, at the 99.73\%
confidence level which at face value, argues for soft cores in
low-mass systems. The annihilation scenario naturally explains
this since the cores are more evident in low-mass systems (see
Fig.~\ref{profiles}).
However, it should be noted \cite{swaters} that the
error bars on the rotation velocity data are probably not a 
complete description of the total uncertainty and that a 
critical reevaluation might lead to a less stringent 
bound on $\alpha$, thus alleviating the discrepancy between 
the observed dwarf rotation curves and CDM predictions.

\begin{figure}[thbp]
\plotone{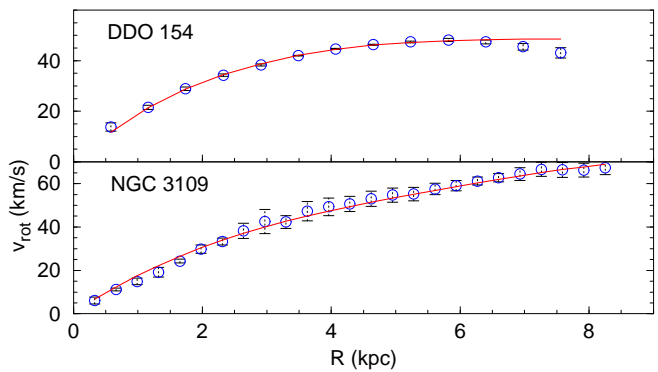}
\caption[caption]{
Rotation curve fits with $\rho_A \simeq 0.2 \msun/\pc^3$.
}
\label{fits}
\end{figure}

To estimate the cross-section required to achieve consistency with
observations, we fit to the two dwarf galaxies identified above with
the halo profile in Eq. \ref{n=0}, a thin stellar disk and the 
observed gas. We have included the effect of finite resolution.
We find that $\rho_A \simeq 0.2\msun/\pc^3$ results in a good fit to 
both (see Fig. \ref{fits}). 
In both cases, the outer parts of the halo (determined by $\rho_s$ and 
$r_s$) are consistent with NFW theory.
To see if structures on the largest scales are consistent with the
annihilation scenario, we turn to strong gravitational lensing of 
background galaxies by clusters.
Tyson \etal \cite{cl0024} model the mass distribution in the cluster
CL 0024, which produces multiple distorted images of a background
galaxy, and find evidence for a compact soft core
(of about $35 h^{-1} \kpc$) in the projected density. 
A similar value for the core radius was inferred earlier by Smail
\etal \cite{cl0024} for CL 0024 and other clusters. X-ray studies
(Bohringer \etal \cite{cl0024}) of CL 0024 are also in agreement 
with the above results. However, we urge caution in interpreting 
these results since the evidence for soft cores in clusters is 
largely based on just one cluster (CL 0024).

A value of $\rho_A = 0.2 \msun/\pc^3$ would produce a core density of
about $0.06 \msun/\pc^3$ in a cluster-sized object. We find that this 
core density is consistent with the surface density reconstruction of
CL 0024 by Tyson \etal \cite{cl0024}. The implied CDM mass within
the arc radius (of $107\,h^{-1}\kpc$) is in agreement with the
quoted value of about $1.66 \times 10^{14} \, h^{-1} \msun$ for the
total mass within the arc radius \cite{highervalues}.

{\parindent0pt\it Discussion.}
The s-wave annihilation scenario with a cross-section of 
$\langle\sigma |v|\rangle=10^{-29}(m/\gev)\rcm^2$
produces a core density of about $1\gev/\rcm^3$ 
over widely different scales. Intriguingly, this seems to 
be consistent with observations.

Apart from cuspy cores \cite{sellwood}, 
simulations of non-interacting CDM also predict a much larger
number of sub-halos for a galactic size halo than the observed number
of galactic satellites \cite{moore99}.
Certainly, the s-wave annihilation scenario has a dramatic effect on
the smallest halos, and this could contribute to their destruction. 
However, further study is required to test this hypothesis. 

Another particle-physics solution in which CDM particles have a large 
cross section for self interaction ($\sigma \sim 10^{-24}\,{\rm cm^2}$ 
has been discussed \cite{spergel}.  This possibility is being tested by 
numerical simulations \cite{SIDM}. However, there are indications that 
self interactions lead to halos that are inconsistent with 
observations \cite{jordi00}.  The jury is still out.

The requirements on a model for annihilating CDM are
stringent, but by no means impossible \cite{riotto}.
They point to a particle beyond those currently being
considered, and therefore, to new physics. 
While it is possible that the solution to the CDM cusp problem will
involve the interpretation of the observations or less exotic astrophysics, 
it is appealing to think that the properties of halo cores may
teach us about the fundamental properties of the CDM particle.

\acknowledgements
We gratefully acknowledge support by the DOE, NASA and NSF at
Chicago, and by the DOE and NASA grant NAG 5-7092 at Fermilab.
We thank M. Valluri for useful discussions and pointers. 
We also thank S. Carroll, D. Eisenstein, S. Hannestad, J. Mohr, 
D. Spergel and L. Widrow.


\begin{thebibliography}{ucsc}
\bibitem{citem} The invariance of the core density with respect to the
mass of the halo has also recently been noted by
C. Firmani \etal, astro-ph/0002376.
\bibitem{dispersion}For simplicity of exposition, we have assumed that the
velocity dispersion is independent of radius.
\bibitem{initial}For simplicity, we have neglected the fact that the 
initial time (when the structure collapses) will depend on halo mass.
\bibitem{highervalues}This is higher than the estimates of
Smail \etal \cite{cl0024} and Kassiola \etal \cite{cl0024}
who calculate a mass of about $1 \times 10^{14} \, h^{-1} \msun$, 
in agreement with the recent work by Broadhurst \etal \cite{cl0024}.
\end{thebibliography}
\end{document}